# Magnetic properties and hydrides precipitation observation of nitrogen doping niobium used for accelerator applications

ZiqinYang[a,*], Xiangyang Lu[b], Yuan He[a], Weiwei Tan[b], Shichun Huang[a], Hao Guo[a]

[a]Institute of Modern Physics, Chinese Academy of Science, Lanzhou 730000, China
[b]State Key Laboratory of Nuclear Physics and Technology, Peking University, Beijing 100871, China

**Abstract**

Nitrogen doping study on niobium (Nb) samples used for the fabrication of the superconducting radio frequency (SRF) cavities has been carried out. The samples' surface treatment was attempted to replicate that of the SRF cavities, which includes heavy electropolishing (EP), nitrogen doping and the subsequent EP with different amounts of material removal. The magnetization curves of both the doped and the un-doped samples have been measured, from which the field of the first flux penetration $H_{ffp}$ and the upper critical field $H_{c2}$ were extracted. The thermodynamic critical field $H_c$, the Ginzburg-Landau parameter κ, the penetration depth λ, the coherence length ξ and the superheating field $H_{sh}$ were calculated from the determined reversible magnetization curves. The $H_{sh}$ of the doped samples is obviously smaller than that of the un-doped samples. Direct observation of hydrides precipitation on both the doped and the un-doped samples has been carried out using Scanning Electron Microscope (SEM) with a cold stand at 80K, which showed that the amount of hydrides precipitation on the doped samples was reduced to varying degrees with different amounts of material removal. Under our doping recipe, the doped sample with subsequent EP material removal of 7-9μm corresponds to the minimum nitrogen concentration that can effectively reduce the amounts of hydrides precipitation

*Keywords:* Niobium, Nitrogen doping, Magnetic properties, Hydrides precipitation

## 1. Introduction

The Nb SRF cavities have been widely used in many particle accelerator projects, which cover the application range from high energy colliders [1-3] to light sources [4-6] and small scale industrial applications [7,8]. The unloaded quality factor Q ($Q_0$) is one of the most important factors of the SRF cavities, which is inversely proportional to the cavities' surface resistance $R_s$. High $Q_0$ can efficiently decrease the cryogenic load of the SRF cavities. The post treatment has an important influence on the cavities' radio frequency (RF) performance. The average $Q_0$ of linac coherent light source (LCLS-II) 9-cell 1.3GHz cavities and cryomodules were set to be exceeding $2.7 \times 10^{10}$ at a gradient of 16MV/m at 2K [9]. It means that the $R_s$ of the cavities is less than 10nΩ, and the standard surface treatment procedures including a combination of chemical treatment like EP, buffered chemical polishing (BCP) and heat treatment cannot meet this requirement. The Nitrogen doping is a new surface treatment technique discovered by A. Grassellino [10], which can systematically improve the $Q_0$ of the Nb SRF cavities up to a factor of about 2-4 compared to the standard surface

This work was supported by Major Research Plan of National Natural Science Foundation of China (91426303) and National Postdoctoral Program for Innovative Talents (BX201700257).
* Corresponding author.
Email address: yzq@impcas.ac.cn

treatment procedures. Cornell University [11] and Jefferson Lab (JLab) [12] have successfully repeated and demonstrated the nitrogen doping experiments on both single cell and 9-cell cavities. On one hand, Cornell University and JLab together with Fermi National Accelerator Laboratory (FNAL) searched for the best nitrogen doping recipe [13] and demonstrated that this process can reliably achieve LCLS-II cavity specification in the production acceptance testing settings [14-16]. Presently, nitrogen doping treatment is being transferred from the prototyping and R&D stage to the production stage [17]. On the other hand, fundamental understanding of the nitrogen doping mechanism is being carried out extensively [13, 18-21], but yet remains unclear.

The theoretical studies of the impact of vortex on the residual surface resistance ($R_{res}$) of the SRF Nb cavities were carried out independently by A. Gurevich and G. Ciovati [22] and M. Checchin [23]. They all predicted that $R_{res}$ has a higher sensitivity to trapped flux with a lower electron mean free path (MFP). Dan Gonnella [24] and M. Martinello [25] studied the sensitivity of $R_{res}$ to trapped flux on both doped and un-doped cavities. Both of their experiments showed that the sensitivity of $R_{res}$ to trapped flux has a bell-shaped trend as a function of the electron MFP, which agrees well with M. Checchin's theory. The doped cavities operating at 1.3 GHz fall exactly in the electron MFP region where the sensitivity is larger than standard cavities. Although the amount of nitrogen diffused into Nb is minimal, the impact of it on the sensitivity of $R_{res}$ to trapped flux is demonstrable.

The Nb hydrides are normal conducting at the typical cavity operating temperature of about 2K. Both previous and present studies [26-27] have shown that the amount of lossy non-superconducting Nb hydrides precipitated on the inner surface of the SRF cavities has a significant influence on the $Q_0$ of the cavities, so the hydrides are an important source of $R_{res}$ in the SRF Nb cavities. Special attention is needed to be paid to the precipitation of Nb hydrides on the surface of Nb samples before and after nitrogen doping treatment.

The aim of the present study is to investigate the effects of nitrogen doping treatment on the magnetic properties and the hydrides precipitation of the Nb samples with different treatments. Magnetic measurements on both doped and un-doped Nb samples were carried out with an SQUID magnetometer (Quantum Design MPMS-XL-7). The observation of the hydrides precipitation on both doped and un-doped Nb samples was carried out by using SEM with a cold stand at 80K. The experimental procedures and results are presented below.

2. Experiments

2.1 Sample preparation

The samples' surface treatment was attempted to replicate that of the SRF cavities, which includes heavy EP, nitrogen doping treatment and the subsequent EP with different amounts of material removal. Vertical test results of N-doped SRF cavities with varying doping parameters showed that the 2/6 recipe proposed by FNAL is the best nitrogen doping recipe [13]. The study in this paper focused on this recipe. The details of sample preparation, nitrogen doping treatment and $800^0C$ heat treatment of the Nb samples can be seen in [21].

After nitrogen doping treatment and $800^0C$ heat treatment, the Nb samples were prepared by subsequent EP with different amounts of material removal to study the effects of different contents of N impurity on the magnetic properties and hydrides precipitation of Nb.

*2.2 Magnetization measurement*

The magnetization was measured by using a commercial SQUID magnetometer (Quantum Design MPMS-XL-7) at Peking University. The weight of Nb samples was between 80-260mg. For the consideration of measurement range, the DC module was used with an accuracy of $1\times10^{-8}$ emu. Mounted in a sample holder, the sample was transported to the sample chamber together with a rigid sample rod during which the sample was cooled from room temperature to 4K at zero field (ZFC). After the temperature was stabilized at 4K, a small scanning field of 50Oe was applied to set the sample being centered. The sample's movement length is 4cm. The magnetic field is homogeneous to 0.05% within this region. The measuring range of DC applied field $H_a$ lies between zero and 5000Oe, including an ascending branch and a descending branch. At each measuring point, the sample was driven through the detection coil and stopped at a number of positions over the specific scan length. The current induced by the magnetic moment of the sample would be detected by the superconducting detection coil, which is a second-order gradiometer to reduce the impact of fluctuations and drift in the magnetic field. So the detection coil can provide a highly accurate measurement of the sample's magnetic moment.

*2.3 Observation of hydrides precipitation*

From the complete equilibrium phase diagram presented in Ref. [28], H atoms randomly distributed over tetrahedral sites in the Nb crystal lattice at room temperature. This is α phase. The solubility limit of this phase extends up to $4\times10^3$ wt ppm of H concentration at room temperature. Nb hydrides cannot precipitate on the surface of SRF Nb cavities at room temperature, of which the H concentration is less than 2 wt ppm. As the temperature is lowered, the hydrogen concentration needed to form the Nb hydrides decreases. At 100K, solubility limit of the ε phase hydride is dramatically reduced to about 5 wt ppm. H in metals has tendency to interact with crystal defects like impurity atoms, grain boundaries, and dislocations due to elastic stresses applied to the lattice [29]. H keeps concentrated near the defects and can even reach Nb hydrides precipitation limit, resulting in the precipitation of non-superconducting Nb hydrides. The diffusion rate of hydrogen between 150 K and 60 K remains quite significant, so that hydrogen can move and accumulate to the critical concentrations at nucleation sites. When the temperature is reduced further below 60 K, the diffusion of hydrogen was slowed down so that hydrogen can no longer accumulate to the hydride centers.

To observe the hydrides precipitation on the surface of SRF Nb cavity, the Nb samples should be kept between 150 K and 60 K. The scanning transmission electron microscopy (STEM) can be used to observe Nb hydrides precipitation [30]. Combined with the electron energy loss spectroscopy (EELS), the atomic scale structure information of Nb hydrides can be obtained. Another experimental method that can be used to observe hydrides precipitation is the cryogenic laser scanning confocal microscopy (CLSCM) [31]. The precipitation of hydrides on Nb surface at low temperature can be observed on the micron scale.

In this study, the observation of hydrides precipitation on both doped and un-doped Nb samples was carried out by using the SEM with a cold stand at 80K. The samples to be observed were adhered to the center disk area on the sample holder by conductive silver adhesives. The disc is surrounded by copper, which enables the samples to be cooled effectively. After the sample holder was carefully fixed on the cold stand, the stand was cooled by the 80K nitrogen gas, which was cooled by liquid nitrogen.

Before the cooling of the samples, a number of observation points were selected randomly and their position coordinates were recorded on each sample in order to realize the switching between different observation points timely and exactly. The electron beam current was about 326μA, the accelerating voltage was 10keV and the working distance was about 10.5cm. The samples were cooled from room temperature to 80K within 10 minutes and remained unchanged at 80K.

### 3. Results

*3.1 Magnetic properties*

Magnetization curves of a doped sample and an un-doped sample after demagnetization correction are shown in Fig. 1, which is typical of type II superconductors. The diffused nitrogen clearly influences the magnetization curve and results in higher $H_{c2}$ and smaller $H_{ffp}$. Detailed analysis will be presented in the following to get the dependence of superconducting parameters on material removal.

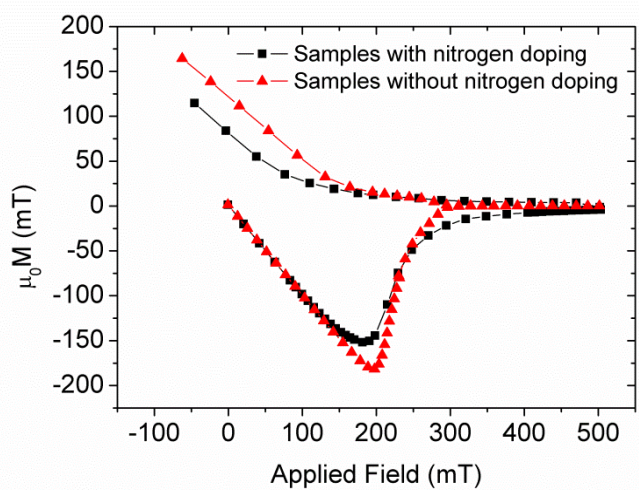

**Fig. 1.** Magnetization curves of a doped sample and an un-doped sample after demagnetization correction are shown for reference.

Flux jumps accompanied by a sudden temperature rise were observed in both the doped and the un-doped samples without subsequent EP removal, which were shown in Fig. 2. Under the same experimental condition, flux jumps were not observed in any samples followed by EP removal. The flux jump phenomenon may be caused by the high concentration of N impurity at the surface layer of the doped sample and the badly polluted surface layer of the heat treated sample. Both of which can provide large amounts of flux pinning centers, resulting in thermos-magnetic instability to cause the sudden redistribution of vortex [32].

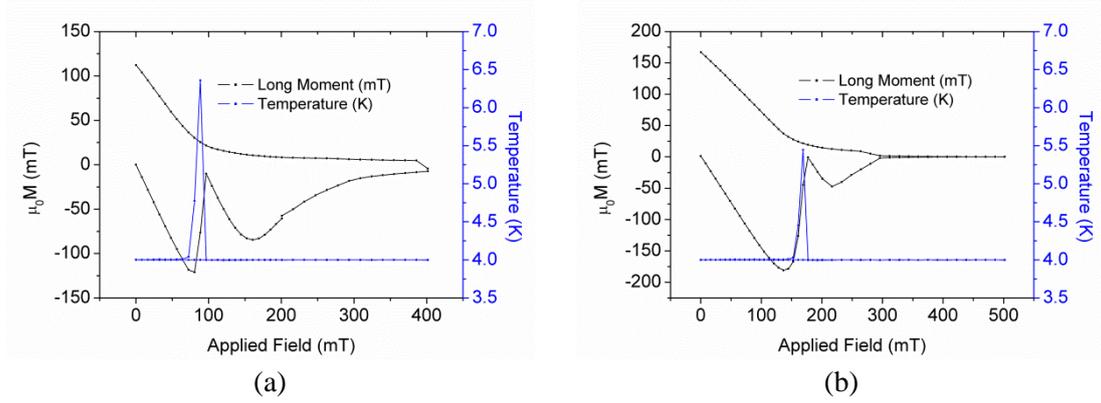

**Fig. 2.** Flux jumps accompanied by sudden temperature rise observed in both doped sample (a) and 800$^0$C heat treated sample (b) without subsequent EP removal.

$H_{ffp}$ is determined as the field at which a deviation starts from the linear magnetization $M(H_a)$ in the Meissner state. $H_{c2}$ is defined as the field at which the magnetization $M(Ha)$ reaches zero.

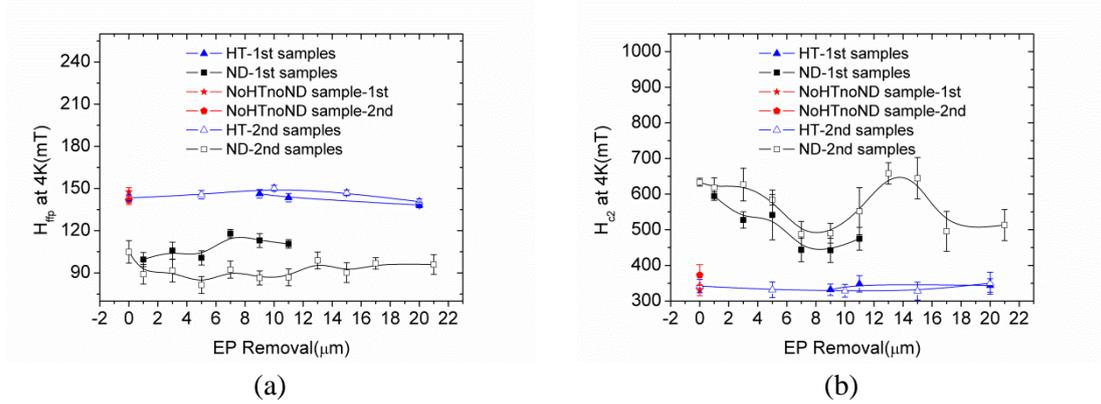

**Fig. 3.** $H_{ffp}$ (a) and $H_{c2}$ (b) of samples with different treatments.

$H_{ffp}$ of the samples with different treatments is shown in Fig. 3(a). $H_{ffp}$ of all samples is determined with the same criteria. $H_{ffp}(0K)$ is extracted from $H_{ffp}(4K)$ by the following empirical relation, which fits well to the $H_{ffp}(T)$ data of cavity-grade Nb [33]:

$$H_{ffp}(T) = H_{ffp}(0)(1 - t^2) \quad (1)$$

where $t = T/T_c$. $H_{ffp}$ of the un-doped samples is about 180mT at 0K. $H_{ffp}$ of all the doped samples is significantly lower compared to the un-doped samples.

$H_{c2}$ of the samples with different treatments is shown in Fig. 3(b). $H_{c2}(0K)$ can be calculated from the temperature dependence deduced from the two-fluid model, which has been proved to be more suitable for the fine grain Nb [33]:

$$H_{c2}(T) = H_{c2}(0)\frac{1-t^2}{1+t^2} \quad (2)$$

$H_{c2}$ of the un-doped samples at 0K is about 480mT. But $H_{c2}$ of the doped samples is significantly higher due to the diffused nitrogen. $H_{c2}$ of the doped samples from both Experiment 1 and Experiment 2 has the same changing trend with the EP removal up to 13μm. Also, in the case of material removal less than 13μm, the doped samples with subsequent EP of 7-9μm have minimum $H_{c2}$. When the subsequent EP removal is larger than 13μm, $H_{c2}$ of the doped samples decreases with the incremental material removal. But both $H_{ffp}$ and $H_{c2}$ of the doped samples with material removal up to 21μm fails to recover to the values of the un-doped samples, which implies the depth of nitrogen doping effect exceeds 21μm under our doping recipe.

A set of superconducting material parameters of the type II superconductors can be deduced from the relationship between $H_{c2}$ and $H_c$.

The Ginzburg-Landau parameter κ can be calculated by:

$$H_{c2} = \sqrt{2}\kappa H_c \tag{3}$$

The penetration depth λ can be calculated by:

$$\lambda(T) = \frac{1}{2 \cdot H_c(T)} \cdot \left(\frac{\hbar \cdot H_{c2}(T)}{e \cdot \mu_0}\right)^{\frac{1}{2}} \tag{4}$$

where $\hbar$ is the Planck's constant divided by $2\pi$.

The coherence length ξ can be deduced from λ(T)/κ. So $H_c$ is the key superconducting parameter for the analysis of the experimental data from magnetization measurements.

For superconductors with high κ value, $H_c$ and κ can be directly estimated from the measured $H_{ffp}$ and $H_{c2}$ [34]. But this method cannot be applied to the SRF Nb because of its low κ value. For superconductors of extremely high purity with low κ value, their magnetization curves are nearly perfect reversible [35] or with little hysteresis [36]. In such conditions, $H_c$ can be directly calculated from the magnetization curves in ascending field:

$$\frac{1}{2}\mu_0 H_c^2 = -\int_0^{H_{c2}} \mu_0 M(H_a) dH_a \tag{5}$$

But the magnetization curves of the SRF Nb samples showed significant hysteresis. The reversible magnetization curves should be reasonably determined to assess the superconducting parameters correctly.

The specific heat and magnetization measurements carried out by R. Radebaugh [37] showed that in the field region of $H_c<H_a<H_{c2}$, the thermodynamically deduced reversible magnetization curve is approximately equal to the average magnetization between increasing and decreasing fields: $M_{rev}=(M_++M_-)/2$, where $M_+$ corresponds to the ascending branch and $M_-$ to the descending branch. This relationship has been verified both experimentally [38] and theoretically [39] and has been used extensively [40] to determine the reversible magnetization curves.

Based on the R. Radebaugh's study, the complete reversible magnetization curve can be divided into four parts. That is the Meissner state part, the part near $H_{ffp}$, the part of intermediate field region (London region) and the part close to $H_{c2}$ (Abrikosov region).

The reversible magnetization curves of the Nb samples in mixed state near $H_{ffp}$ can be deduced by [41]:

$$M = \frac{2\phi_0}{\sqrt{3}\mu_0\lambda^2}\left(\ln\frac{3\phi_0}{4\pi\mu_0\lambda^2(H_a-H_{ffp})}\right)^{-2} - H_a \tag{6}$$

Where $\phi_0 = h/2e = 2.0678 \times 10^{-15}$(Wb) is the quantum flux. This relationship reflects the typical physical characteristics of the ideal type II superconductors that it exhibits infinite magnetic permeability near $H_{ffp}$.

In the intermediate field region of $H_{ffp}<H_a<<H_{c2}$, the London model [42] has provided the only detailed phenomenological description for type-II superconductors. The London model predicts a linear dependence of M on $\ln H_a$ in London region and agrees well with experiments [43]. The κ values of the doped Nb samples lie between 2 to 3. The reversible magnetization curves of the Nb samples in the London region can be approximately determined based on London model by the fit:

$$M = -a - b\ln\frac{c}{H_a} \quad (7)$$

where a, b and c are the fitting parameters.

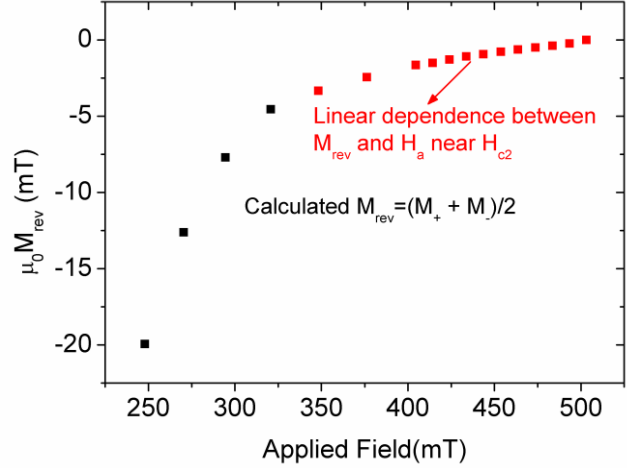

**Fig. 4.** Linear dependence between $M_{rev}$ and $H_a$ near $H_{c2}$ of a doped sample.

When $H_a$ is close to $H_{c2}$, the London model is no longer applicable due to the depression of the order parameter to zero at the vortex centers. The magnetization of the ideal type II superconductors varies linearly with $H_a$ near $H_{c2}$ [42]. The calculated $M_{rev}=(M_+ + M_-)/2$ of both the doped and the un-doped samples near $H_{c2}$ show good linear behavior as shown in Fig. 4. So the linear region of the calculated $M_{rev}$-$H_a$ curve near $H_{c2}$ can be considered as the reversible magnetization curve of the Abrikosov region. Also, the linear behavior of $M_{rev}$ with $H_a$ close to $H_{c2}$ can be used to identify the demarcation point between the London region and the Abrikosov region.

The reversible magnetization curve of the London region covering from $H_c$(~200mT for the SRF Nb) to the demarcation point can be fit based on the London model, which is shown in Fig. 5.

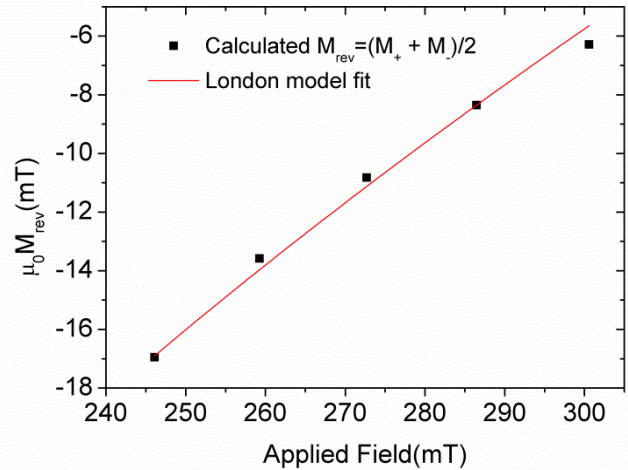

**Fig. 5.** London model fitting of a doped sample is shown as a reference.

The reversible magnetization curve near $H_{ffp}$ can be calculated by equation (6). Combining the three regions of reversible magnetization curve between $H_{ffp}$ and $H_{c2}$, together with the part in Meissner state from zero to $H_{ffp}$, the complete reversible magnetization curve can be determined, which is shown in Fig. 6. $H_c$ can thus be calculated from the determined reversible magnetization curve by equation (5).

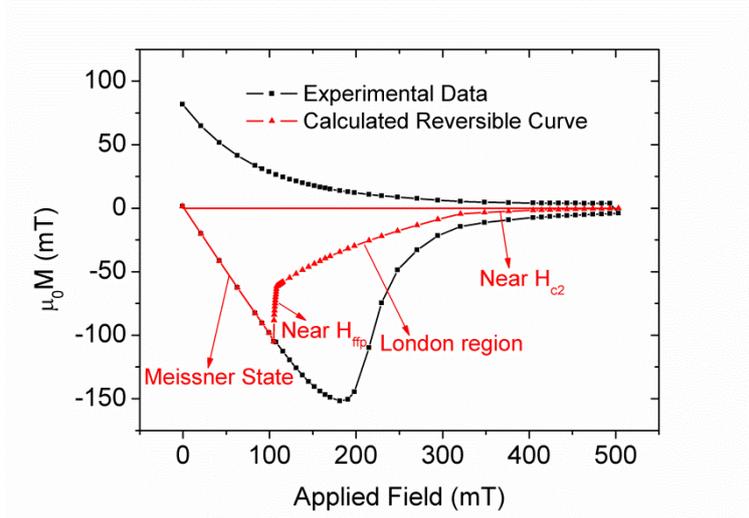

**Fig. 6.** The determined complete reversible magnetization curve of a doped sample that reflects the typical physical characteristics of the ideal type II superconductors.

$H_c$ of the Nb samples with different treatments is listed in Table 1. The $H_c$ averaged over the un-doped samples is 208±10mT, which agrees with the value $H_c(0) = 199\pm1$mT reported in Ref. [35]. $H_c$ of the doped samples is slightly smaller.

**Table 1** $H_c$ of samples with different treatments.

| Experiment 1 | | Experiment 2 | |
| --- | --- | --- | --- |
| **Samples** | $H_c$ at 0K/mT | **Samples** | $H_c$ at 0K/mT |
| ND-1st-1μm[1] | 190±1 | ND-2nd-0μm[2] | 189±12 |
| ND-1st-3μm | 189±1 | ND-2nd-1μm | 180±1 |
| ND-1st-5μm | 192±1 | ND-2nd-3μm | 185±4 |
| ND-1st-7μm | 193±1 | ND-2nd-5μm | 181±1 |
| ND-1st-9μm | 187±2 | ND-2nd-7μm | 187±9 |
| ND-1st-11μm | 183±2 | ND-2nd-9μm | 185±1 |
| HT-1st-9μm | 214±6 | ND-2nd-11μm | 185±8 |
| HT-1st-11μm | 210±2 | ND-2nd-13μm | 172±1 |
| HT-1st-20μm | 201±1 | ND-2nd-15μm | 181±1 |
| | | ND-2nd-17μm | 182±1 |
| | | ND-2nd-21μm | 185±1 |
| | | NoHTnoND[4-1] | 195±4 |
| | | NoHTnoND[4-2] | 204±6 |
| | | HT-2nd-0μm[3] | 204±6 |
| | | HT-2nd-5μm | 207±1 |
| | | HT-2nd-10μm | 219±1 |
| | | HT-2nd-15μm | 200±2 |
| | | HT-2nd-20um | 229±2 |

[1] N-doped sample from experiment 1 with 1μm subsequent material removal

[2] N-doped sample from experiment 2 with 0μm subsequent material removal

[3] 800$^0$C heat treated sample from experiment 2 with 0μm subsequent material removal

[4-1] First test of the sample after EP 150μm without neither nitrogen doping treatment nor 800$^0$C heat treatment.

[4-2] Second test of the sample after EP 150μm without neither nitrogen doping treatment nor 800$^0$C heat treatment.

The calculated superconducting parameters such as κ, ξ and λ can be seen in Fig. 7. The κ and λ of the doped samples are higher than that of the un-doped samples, while the ξ of the doped samples is lower.

H. Ullmaier [44] studied the pinning of flux lines in Nb due to statistically distributed vacancies and interstitials (Frenkel pairs), which were produced by low temperature 3MeV electron irradiation. H. Ullmaier's study showed that when the mean distance between the point defects is smaller than the coherence length, the fluctuations in the defect density are responsible for the observed pinning effects. In the case of nitrogen doping, the N concentration of the Nb surface layer lies roughly between $10^{20}$ to $10^{18}$ atoms/cm$^3$ [45]. The corresponding mean distance between the N impurities falls exactly within the research region of H. Ullmaier's study. On the other hand, E. V. Thuneberg et al [46] theoretically calculated the elementary flux pinning behavior between flux lines and small size defects (typically point defects) based on a strong pinning mechanism brought about by quasiparticle scattering off the pinning centers. The physical basis of this pinning mechanism is a nonlocal effect that a scattering center has on Cooper pairs in its immediate environment. A scattering center helps a superconductor to sustain deformations of the order parameter up to distances on the order of the zero-temperature coherence length. Hence it is energetically advantageous for a region where the order parameter varies strongly, e.g., a vortex core, to coincide with a scattering center. So although the dimension of a single N atom is much smaller than the coherence length of Nb, such a concentration of N impurity distribution may be responsible to the observed $H_{c2}$, κ, ξ and λ changes.

When $H_a > H_{ffp}$, the vortex flux favors to enter the bulk material region in the manner of quantum magnetic flux. But the entry of the vortex flux lines will be obstructed by the pinning forces (grain boundaries, dislocations, and the collective behavior of N impurities, etc.) until $H_a$ continues to increase to overcome the pinning forces. After the flux vortex lines have just entered the superconductor, the distance between the flux vortex lines is large and the interaction between the flux vortex lines is small, so more flux vortex lines can enter into the superconductor relatively quickly. However, as $H_a$ is further enhanced, the distance between the flux vortex lines becomes smaller, creating resistance to the entry of the new flux vortex lines due to the repulsive Lorenz force. The pinning forces also prevent the flux lines from moving deeper into the superconductor. The representation to the magnetization curve is that, when the magnetization reaches the minimum, the magnetization changes rapidly with $H_a$ initially, and then the magnetization increases with $H_a$ more and more slowly until $H_a$ increases to $H_{c2}$ with the transition to the normal state. For an N-doped Nb sample with higher concentration of N impurity, the obstruction to the continuously entry of vortex flux lines into the bulk material offered by the pinning forces will be larger. Until when $H_a$ is increased to a larger value, sufficient flux vortex lines can enter the Nb. The superconducting parameters such as κ, ξ, λ were determined from $H_{c2}$ and $H_c$. According to the calculation of $H_c$, the nitrogen doping treatment did not make much difference to $H_c$. As a result, the changing trends of κ, ξ and λ with EP removal coincide with that of the $H_{c2}$, which is originated from the varying concentration of the doped nitrogen. Thus, for our specific study in this paper, the superconducting parameters κ, ξ and λ were calculated from the magnetization measurements and they may reflect the magnetic properties of the N-doped Nb samples.

These three parameters, together with $H_{c2}$, reflect a common phenomenon that the observed pinning effects of the doped samples are more significant. The κ and λ of the doped samples have local minimum values between EP material removals of 7μm to 9μm, which is also the location of the local maximum values of ξ. This means the doped samples with subsequent EP material removals of 7-9μm have least pinning effects in the case of material removal less than 13μm under our doping recipe. When the subsequent material removal is less than 7μm, the observed pinning effects decrease in the incremental material removal. On the contrary, when the subsequent EP material removal is larger than 9μm but less than 13μm, the observed pinning effects increase with the sustained reduction of N impurity. When the subsequent EP material removal is larger than 13μm, with the further reduction of N concentration, the pinning effects is gradually weakened with the incremental EP material removal. But until the EP material removal of the doped sample reached 21μm, the κ, ξ and λ still did not recover to the level of the un-doped samples. And this is in consistent with the phenomenon reflected by the $H_{ffp}$ and $H_{c2}$ changing trends.

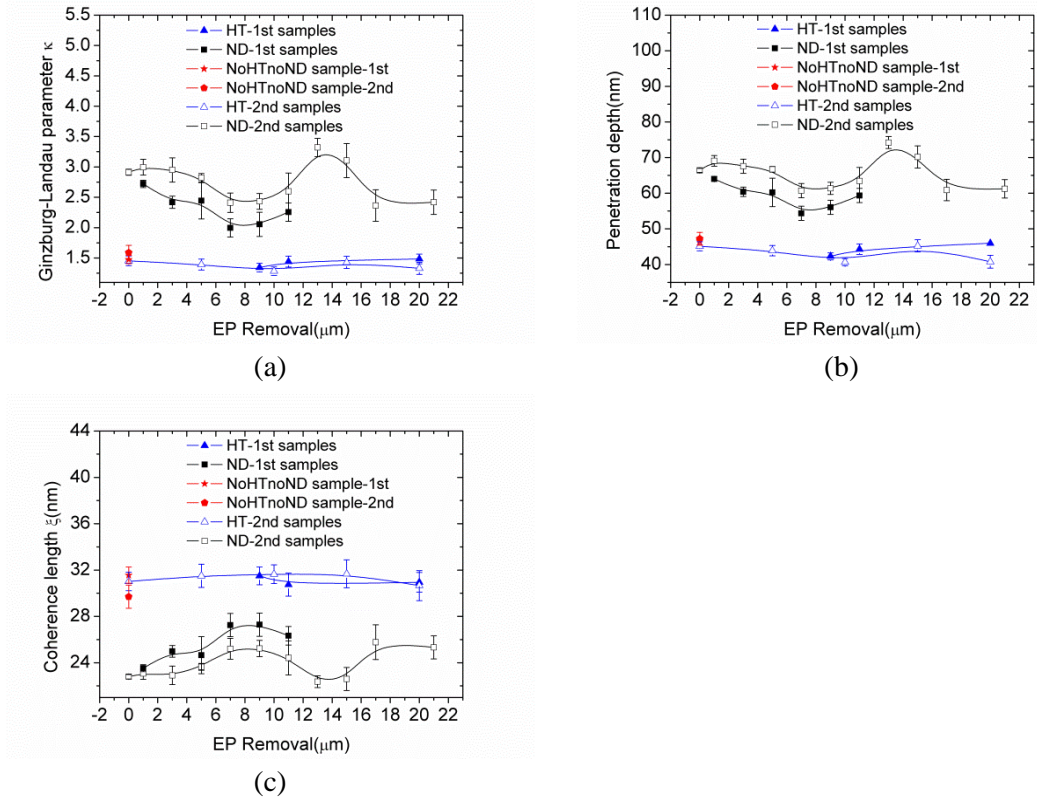

**Fig. 7.** Superconducting parameters κ(a), λ(b) and ξ(c) of samples with different treatments.

Because of the entropy discontinuity and required nucleation centers during the first order phase transition taking place in the presence of an external field, there is a possibility for the superconducting state to persist as a metastable state above $H_{ffp}$ up to $H_{sh}$ [47], which has been verified by the high pulsed power measurements on both the N-doped cavities [48] and the cavities with standard treatment [49]. Therefore, $H_{sh}$ is the fundamental limit to the accelerating field of the SRF cavities.

$H_{sh}$ of the SRF Nb can be calculated by using a corrected formulation valid for k≥1 [50]:

$$H_{sh} = \sqrt{2}\left(\frac{\sqrt{10}}{6} + \frac{0.3852}{\sqrt{\kappa}}\right)H_c \qquad (8)$$

Substituting the calculated $H_c$ and κ value into equation (8), $H_{sh}$ can be obtained. $H_{sh}$ of the samples with different treatments at 0K is shown in Fig. 8.

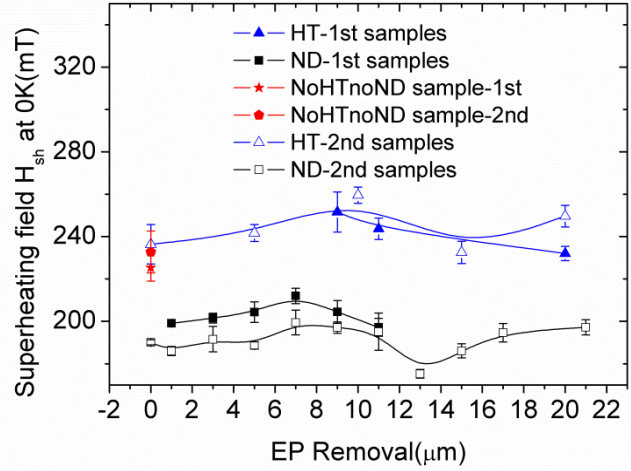

**Fig. 8.** $H_{sh}$ of samples with different treatments.

$H_{sh}$ averaged over the un-doped samples is 240±10mT at 0K, which is in consistent with the estimated value of 240mT [47, 49]. The largest $H_{sh}$ value of the doped samples in Experiment 1 and Experiment 2 at 0K is 211±4mT and 199±5mT, respectively. $H_{sh}$ of the doped samples is obviously smaller than that of the un-doped samples. The dependence of $H_{sh}$ on EP material removal implies the possible changing trends of $E_{acc}$ along with the material removal for the defect free N-doped Nb cavities.

*3.2 Hydrides precipitation*

The surface morphology changes of the NoHTnoND sample at the same observation position and at different temperature are shown in Fig. 9 for the verification of the Nb hydrides. The white island bumps precipitated at 80K disappeared when the temperature was raised up to 206K. Combined with the previous study [31] and Nb-H phase diagram, it can be verified that the observed white island bumps are the precipitated Nb hydrides. Observations showed that as soon as the temperature was reduced to 80K, obvious hydrides have been precipitated on the surface of the NoHTnoND sample. When the NoHTnoND sample was kept at 80K for 50 minutes, the size and amounts of hydrides did not change substantially over time. The defects shown in the figures should be the mechanical scratches left after EP material removal of 150μm. Furthermore, hydrides are more easily precipitated at the defect sites on the sample surface. The small-roughness and defect-free inner surface of the SRF cavities may be beneficial to its avoidance of hydrides precipitation.

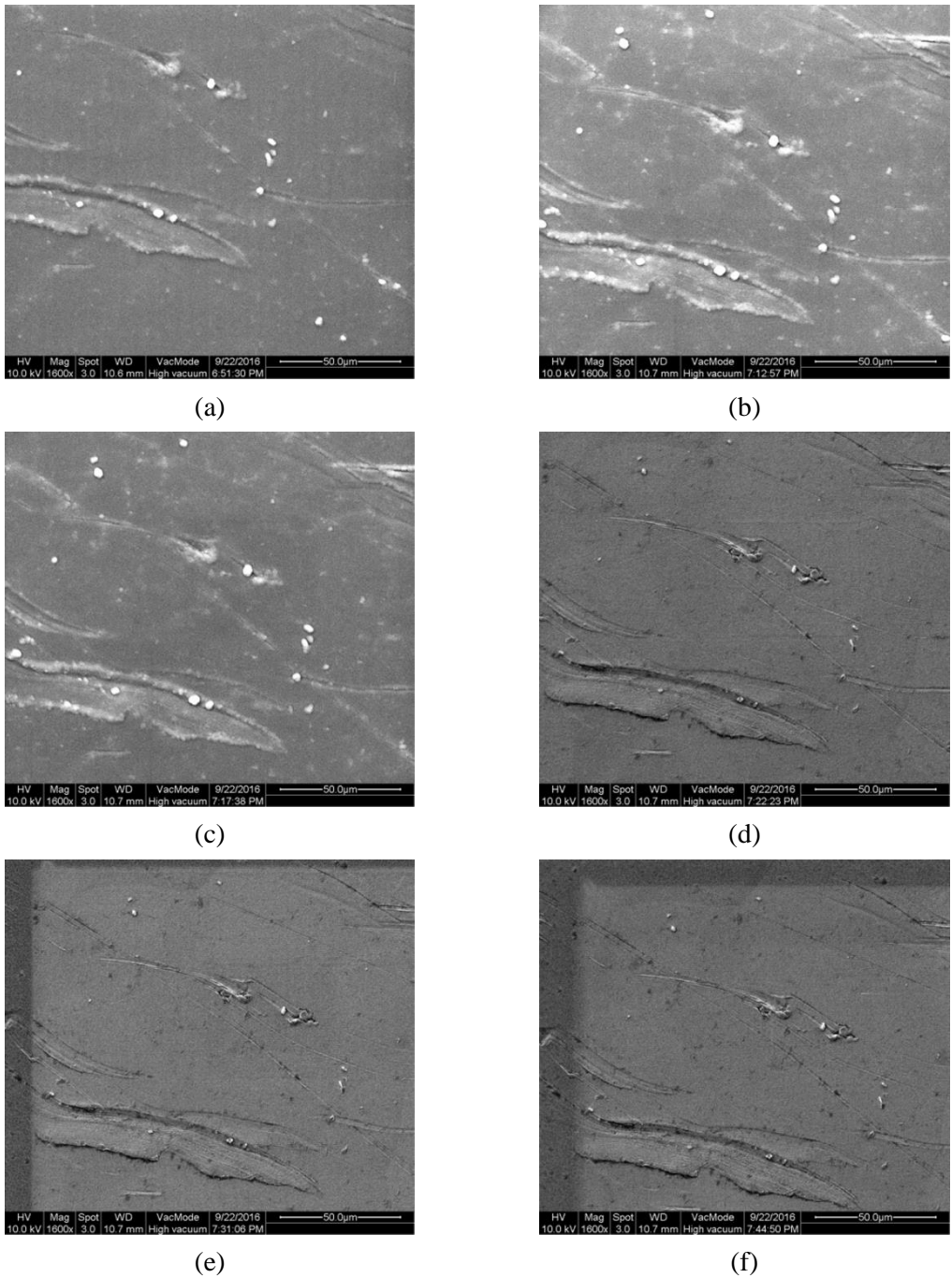

**Fig. 9.** At the same observation position, the surface morphology of the NoHTnoND sample at temperature: (a) 80K, (b) 190K, (c) 198K, (d) 206K, (e) 250K, (f) 298K.

Comparison of the specific precipitation of Nb hydrides at random observation positions in each sample is shown in Fig. 10. The surface of the ND-2$^{nd}$-0μm sample has corrosion pits caused by polishing of sand paper.

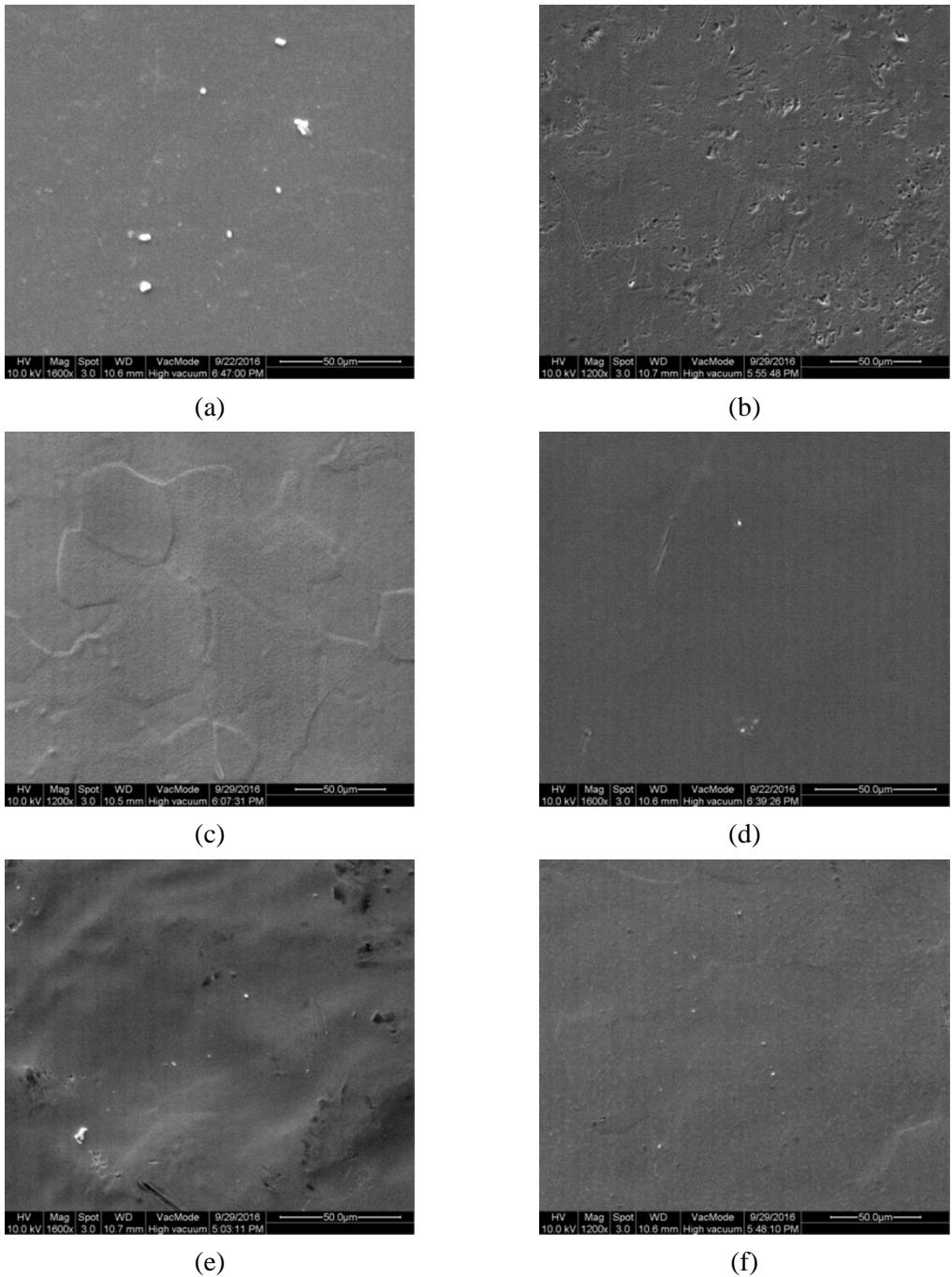

**Fig. 10.** Comparison of the specific precipitation of the Nb hydrides at random observation positions in each sample: (a) the NoHTnoND sample, (b) the ND-2$^{nd}$-0μm sample, (c) the ND-2$^{nd}$-7μm sample, (d) the ND-2$^{nd}$-17μm sample, (e) the ND-2$^{nd}$-21μm sample, (f) the HT-2$^{nd}$-20um sample.

The largest number and largest size of the Nb hydrides were precipitated on the surface of the NoHTnoND sample. The characteristic size of the island bump shaped hydrides on the surface of the NoHTnoND sample was about 3μm×3μm. The HT-2$^{nd}$-20μm sample showed a significant different precipitation of hydrides compared to the NoHTnoND sample. There was also a considerable amount of hydrides precipitated on the surface of the HT-2$^{nd}$-20μm sample, but the size of the hydrides was about 1.5μm×1.5μm. This implies that the H content in SRF Nb can be effectively reduced by $800^0$C heat treatment and therefore to prevent Q disease.

No Nb hydrides were observed at any observation position on the surface of both the ND-$2^{nd}$-0μm sample and the ND-$2^{nd}$-7μm sample. However, the Nb hydrides began to precipitate on the surface of the ND-$2^{nd}$-17μm sample again. The amounts of the Nb hydrides precipitated on the surface of the ND-$2^{nd}$-21μm sample were further enhanced with the sustained reduction of N concentration. The characteristic size of the hydrides precipitated on the surface of the ND-$2^{nd}$-17μm sample and the ND-$2^{nd}$-21μm sample was comparable with that of the HT-$2^{nd}$-20μm sample. But the amounts of the Nb hydrides did not recover to the value of the HT-$2^{nd}$-20μm sample, which implies that the depth of nitrogen doping effect exceeds 21μm under our doping recipe. The observations directly showed that the amounts of the hydrides precipitation can be prevented or retarded to varying degrees with different amounts of material removal, which corresponds to the varying concentrations of the doped nitrogen.

4. Discussion

The residual resistance is the saturation value of the surface resistance tending at very low temperatures (<1.5K), which consists of the contribution from several aspects. Among which the trapped flux is an important source. The unloaded quality factor of the SRF Nb cavities can be systematically improved up to a factor of about 2-4 by appropriate nitrogen doping treatment. But the N-doped cavities show higher dissipation per unit of magnetic flux trapped than the cavities with the standard treatment. This is the motivation of a detailed study of the effect of nitrogen doping treatment on the magnetic properties of the Nb samples used for the fabrication of the SRF cavities. The positions of the flux jumps are both immediately after the minimum magnetization for the N-doped and the $800^0$C heat treated samples without subsequent EP material removal. This may indicate that the pinning mechanism is same for both samples, which may be the high concentration of impurities in the surface layers of the samples. The doped samples have lower $H_{ffp}$ and higher $H_{c2}$ and this is consistent with the physical image. Based on the DC magnetization measurements, the superconducting material parameters were calculated from the determined reversible magnetization curves. The value of κ, ξ, λ and $H_{c2}$ indicates that the observed pinning effects in the doped samples are more significant. This implies that although much more efficient flux expulsion can be achieved by fast cooling down through $T_c$ [51] and no significant effect of surface treatments on flux trapping was observed under the existing experimental conditions by fluxgate magnetometers attached to the middle of the cavities cell in S. Posen's study [52], special attention is still needed to be paid to the ambient magnetic field shielding. The largest $H_{sh}$ values of the doped samples of Experiment 1 and Experiment 2 at 0K is 211±4mT and 199±5mT, respectively. For the typical elliptical cavity geometry, $H_p/E_{acc}$=4.2mT/MV/m leads to $E_{acc}$~48MV/m as the fundamental limit to the N-doped and defect free Nb cavities under our doping recipe. Flux vortex may penetrate into the cavities at field-enhanced defects, which may cause a smaller breakdown field.

Depth profiles of N in the doped samples by SIMS measurements [53] revealed that the concentration of N decreases with incremental EP material removal, resulting in fewer and fewer pinning centers. However, the changing trends of κ, ξ, λ and $H_{c2}$ have a common inflection point at the same EP material removal of 7-9μm, which implies that, except for the pinning effects possibly due to the N impurity distribution, there may exist another kinds of pinning mechanism that changes with the subsequent EP material removal. Considering the significant influence of the amounts of precipitated lossy non-superconducting Nb hydrides on the $Q_0$ of the SRF cavities, together with the previous studies that N impurity can reduce the hydrogen diffusion coefficient in Nb at low temperatures by both quasielastic neutron scattering method [54] and Gorsky-effect measurements [55], the Nb hydrides were suspected to be responsible for the pinning mechanism related to the concentration of N impurity. This is the specific need to study the hydrides precipitation of the N-doped SRF Nb and the research results have indeed proved this conjecture. For the N-doped SRF Nb under our doping recipe, when the EP material removal is less than 7μm, the amounts of the precipitated hydrides can be effectively lowered due to the reduction of the hydrogen diffusion coefficient. However, the relatively high concentration of N impurity leads to relatively pronounced pinning effect. When the EP material removal reaches 7-9μm, the N concentration is remarkably reduced but it is still sufficient to effectively prevent the precipitation of hydrides, both of which results in the least noticeable pinning effects. When the EP material removal is larger than 9μm, with the sustained reduction of N concentration, the reduction of the diffusion coefficient of hydrogen in Nb is insufficient to effectively avoid the precipitation of hydrides. More hydrides emerged with the continuous increase of EP material removal. In this case, the pinning effect becomes more pronounced again due to the trap of magnetic flux by the lossy non-superconducting hydrides. As the amounts of EP material removal exceeds 13μm, in spite of the precipitation of hydrides, the pinning effects becomes less and less obvious due to the further reduction of N impurities. Nevertheless, until to the EP material removal of 21μm, both the pinning effects and the amounts of hydrides did not recover to the level of the un-doped Nb samples.

5. Conclusion

A detailed study and analysis of the magnetic properties of the N-doped SRF Nb has been carried out by DC magnetization measurements. The diffused nitrogen results in significantly lower $H_{ffp}$ and noticeably higher $H_{c2}$. More significant pinning effects were observed in the doped Nb samples, which indicate that the SRF Nb cavities after nitrogen doping treatment need more attention to be paid to the efficient flux expulsion and the more stringent ambient magnetic field shielding. According to the complete reversible magnetization curve determined in four intervals, the superconducting parameters such as κ, ξ, λ and $H_{sh}$ of each sample were calculated. This provides a convenience for exploring the effects of different post-treatments on the superconductivity of the SRF Nb. $H_{sh}$ of the doped samples is obviously smaller than that of the un-doped samples, which may be a possible reason for the reduction of the achievable $E_{acc}$ in the N-doped SRF Nb cavities.

Direct observation of hydrides precipitation on both the doped and the un-doped Nb samples at 80K has been carried out by using SEM with a Cold Stand. The amount of hydrides of the doped samples was reduced to varying degrees with different amounts of material removal, which corresponds to the varying concentrations of N impurity. Under our doping recipe, the doped sample with subsequent material removal of 7-9μm corresponds to the minimum nitrogen concentration that can effectively reduce the amounts of hydrides precipitation, and the doped sample of this condition has the least noticeable pinning effects. This study helps to understand the mechanism of the residual resistance of the N-doped SRF cavities with the subsequent EP material removal. From the analysis of the experimental data in this study, the N-H interaction may be an important factor to reveal the physical mechanism of nitrogen doping phenomenon.

## Acknowledgments

The authors are grateful to L. Lin (Peking University) for his help in EP. The authors would also acknowledge Dr. Y. Zhang (Peking University) for the help in magnetization measurements and Dr. L. Chen for the help in hydrides observation. This work is supported by Major Research Plan of National Natural Science Foundation of China (91426303) and National Postdoctoral Program for Innovative Talents (BX201700257).